\documentclass[12pt]{article}
\usepackage{amsmath}
\usepackage{amssymb}
\usepackage{cite}
\newcommand{\Was}{W\c as}
\newcommand{\eqn}[1]{(\ref{#1})}

\newcommand{\rQED}{{\rm QED}}
\newcommand{\rQCD}{{\rm QCD}}
\newcommand{\rQCED}{{\rm QCED}}
\newcommand{\rYFS}{{\rm YFS}}
\newcommand{\rBorn}{{\rm Born}}
\begin{document}
\title{QED$\otimes$QCD Resummation and Shower/ME Matching for LHC Physics%
\thanks{Presented by S.A.Y. at the Cracow Epiphany Conference on Precision
Physics and Monte Carlos for LHC, 4 -- 6 January, 2007.}%
}
\author{{\large B.F.L. Ward and S.A. Yost}\\
Department of Physics, Baylor University, Waco, TX, USA\\\qquad \\
BU-HEPP-07/02
}
\maketitle

\begin{abstract}
We present the theory of QED$\otimes$QCD resummation and its interplay 
with shower/matrix element matching in precision LHC physics scenarios. 
We illustrate the theory using single heavy gauge
boson production at hadron colliders.
\end{abstract}

PACS numbers: 12.38.Cy, 12.15.Lk, 11.25.Db

\section{Introduction}
In the imminent LHC environment, where one expects to have an experimental 
luminosity precision tag at the level of 2\%, \cite{lhclum} the 
requirement for the theoretical precision tag on the corresponding luminosity
processes, such as single $W, Z$ production with the subsequent decay into 
light lepton pairs, should be at the 0.67\% level in order not to compromise,
unnecessarily, the over-all precision of the respective LHC luminosity 
determinations. This dictates that multiple gluon and photon radiative 
effects must be controlled at the stated precision. The theory of 
$QED\otimes QCD$ exponentiation~\cite{qced} allows for the simultaneous 
resummation of multiple gluon and multiple photon radiative effects in 
LHC physics processes, to be realized ultimately by MC
methods on an event-by-event basis in the presence of parton showers,
in a framework which allows us to systematically improve the accuracy
of the calculations without double-counting of effects, in principle
to all orders in both $\alpha_s$ and $\alpha$. Such a theoretical
framework opens the way to the desired theoretical precision tag on the 
LHC luminosity processes.

Our starting point for the new QED$\otimes$QCD resummation theory~\cite{qced}
is the QCD resummation theory presented in Ref.~\cite{qcdref}. This 
resummation is an exact rearrangement of the QCD perturbative series based
on the $N=1$ term in the exponent in the formal proof of 
exponentiation in non-Abelian gauge theories {\it in the eikonal approximation},
as given in Ref.~\cite{gatherall}. This exponential is augmented with a sum of 
residuals which take into account the remaining contributions to the 
perturbative series exactly to all orders in $\alpha_s$.\footnote{If desired, 
our overall expoential factor can be made to include all of the terms in the 
exponent in Ref.~\cite{gatherall}, in principle.} We therefore have an exact 
result whereas the resummation theory in Ref.~\cite{gatherall} and
those in Refs.~\cite{sterman,cat-tren,berger} are approximate. 
Recently, an alternative resummation theory, the soft-collinear effective 
theory(SCET)~\cite{bauer}, has been developed to treat double resummation 
of soft and collinear effects. Since we have an exact re-arrangement of the 
perturbative series, we could introduce the results from 
Refs.~\cite{sterman,cat-tren,berger,bauer} into our representation as well. 
Such introductions will appear elsewhere.

The need for the extension of the QCD resummation theory to QED$\otimes$ QCD
resummation was already suggested by the results in 
Refs.~\cite{cern2000,spies,james1,roth,james2,blumlein}, where it was shown 
that in the evolution of the structure functions the inclusion of the QED 
contributions leads to effects at the level of $\sim 0.3\%$, already almost 
half of the error budget discussed above. We will find similar size effects 
from the threshold region of heavy gauge boson production. All of these must 
be taken into account if one wants $\sim 1.0\%$ for the theoretical precision 
tag.

The discussion is organized as follows. In Section 2, we review the extension 
of the YFS theory to an exact resummation theory for QCD. Section 3 presents 
the further extension to QED$\otimes$QCD. Section 4 contains the application 
to heavy gauge boson production with the attendant discussion of shower/ME 
matching. Section 5 contains some concluding remarks.

\section{Extension of YFS Theory to QCD} 
We consider a parton-level single heavy boson production process such 
as $q+\bar q'\rightarrow V+n(g)+X\rightarrow
\bar{\ell} \ell'+n(g)+X$,
where $V=W^\pm,Z$, and $\ell = e,\mu,~\ell'=\nu_e,\nu_\mu ( e,\mu )$
respectively for $V=W^+ (Z)$, and  
$\ell = \nu_e,\nu_\mu,~\ell'= e,\mu$ respectively for $V = W^-$.
It has been established~\cite{qcdref} that the cross section may
be expressed as 
\begin{equation}
\begin{split}
d\hat\sigma_{\rm exp} = \sum_n d\hat\sigma^n 
&= e^{\rm SUM_{IR}(QCD)}\sum_{n=0}^\infty\int\prod_{j=1}^n\frac{d^3 k_j}{k_j}\\
& \int\frac{d^4y}{(2\pi)^4}e^{iy\cdot(p_1+p_2-q_1-q_2-\sum k_j)+ D_\rQCD}\\
&\times \tilde{\bar\beta}_n(k_1,\ldots,k_n)\frac{d^3p_2}{p_2^{\,0}}
	\frac{d^3q_2}{q_2^{\,0}}
\label{qcd}
\end{split}
\end{equation}
where gluon residuals 
$\tilde{\bar\beta}_n(k_1,\ldots,k_n)$, defined by Ref.~\cite{qcdref}, 
are free of all infrared divergences to all 
orders in $\alpha_s(Q)$. The functions ${\rm SUM}_{IR}(\rQCD)$ and $D_\rQCD$, 
together with the basic infrared functions 
$B^{nls}_{\rQCD}$, ${\tilde B}^{nls}_{\rQCD}$, and ${\tilde S}^{nls}_{\rQCD}$ 
are specified in Ref.~\cite{qcdref}.  
We call attention to the essential compensation between
the left over genuine non-Abelian IR virtual and real singularities
between the phase space integrals $\int d{\rm Ph}\;\bar\beta_n$ and 
$\int d{\rm Ph}\;\bar\beta_{n+1}$ 
that really allows us to isolate $\tilde{\bar\beta}_j$ and distinguishes
QCD from QED, where no such compensation occurs.
The result in \eqn{qcd} has been realized by Monte
Carlo methods~\cite{qcdref}.
See also Refs.~\cite{van1,van2,anas} for exact ${\cal O}(\alpha_s^2)$
and Refs.~\cite{baurall,ditt,russ} for exact ${\cal O}(\alpha)$
results on the heavy gauge boson production processes which we discuss here.

Apparently, we can not emphasize too much the exactness of \eqn{qcd}.
Some confusion seems to exist because it does not show explicitly an ordered
exponential operator for an appropriate ordering prescription, path-ordered,
time-ordered, etc. The essential point is that, in \eqn{qcd}, we have 
evaluated the matrix elements of these operators and written the result in 
terms of the over-all exponent shown therein and the residuals 
$\tilde{\bar\beta}_j$.  This allows us to maintain exactness 
to all orders in $\alpha_s$.

\section{QED$\otimes$QCD Resummation Theory}

The new $QED\otimes QCD$ theory 
is obtained by simultaneously exponentiating the large 
IR terms in QCD and the exact IR divergent terms in QED, so that
we arrive at the new result
\begin{equation}
\begin{split}
d\hat\sigma_{\rm exp} = &\ e^{\rm SUM_{IR}(QCED)}
   \sum_{{n,m}=0}^\infty\int\prod_{j_1=1}^n\frac{d^3k_{j_1}}{k_{j_1}} 
\prod_{j_2=1}^m\frac{d^3{k'}_{j_2}}{{k'}_{j_2}}\\
& \int\frac{d^4y}{(2\pi)^4}
    e^{iy\cdot(p_1+q_1-p_2-q_2-\sum k_{j_1}-
	\sum {k'}_{j_2})+ D_\rQCED} \\
    & \times \tilde{\bar\beta}_{n,m}(k_1,\ldots,k_n;k'_1,\ldots,k'_m)
	\frac{d^3p_2}{p_2^{\,0}}\frac{d^3q_2}{q_2^{\,0}},
\end{split}
\label{qced}
\end{equation}
where the new YFS~\cite{yfs,yfs1} residuals, 
$\tilde{\bar\beta}_{n,m}$$(k_1,\ldots,k_n;$$k'_1,\ldots,k'_m)$, 
with $n$ hard gluons and $m$ hard photons,
defined in Ref.~\cite{qced}, represent the successive application
of the YFS expansion first for QCD and subsequently for QED. 

The functions ${\rm SUM_{IR}(QCED)}$, $D_\rQCED$ are determined
from their QCD analogs ${\rm SUM_{IR}(QCD)},D_\rQCD$ via the
substitutions
\begin{eqnarray}
B^{nls}_{\rQCD} &\rightarrow B^{nls}_{\rQCD}+B^{nls}_{\rQED} 
	&\equiv B^{nls}_{\rQCED}, \nonumber\\
{\tilde B}^{nls}_{\rQCD} &\rightarrow {\tilde B}^{nls}_{\rQCD} +
	{\tilde B}^{nls}_{\rQED} &\equiv {\tilde B}^{nls}_{\rQCED}, \\
{\tilde S}^{nls}_{\rQCD} &\rightarrow {\tilde S}^{nls}_{\rQCD} 
	+ {\tilde S}^{nls}_{\rQED} &\equiv {\tilde S}^{nls}_{\rQCED}\nonumber
\label{irsub}
\end{eqnarray}
everywhere in expressions for the
latter functions given in Ref.~\cite{qcdref}. We stress that if desired
the exponent corresponding the $N^{\rm th}$ Gatherall exponent for $N > 1$ can 
be systematically included in the QCD exponents ${\rm SUM_{IR}(QCD)}$, $D_\rQCD$
if desired, with a corresponding change in the respective residuals
$\tilde{\bar\beta}_{n,m}$$(k_1,\ldots,k_n;$ $k'_1,\ldots,k'_m)$.
The residuals $\tilde{\bar\beta}_{n,m}$$(k_1,\ldots,k_n;$ $k'_1,\ldots,k'_m)$ 
are free of all infrared singularities,
and the result in \eqn{qced} is a representation that is exact
and that can therefore be used to make contact with parton shower 
MC's without double counting or the unnecessary averaging of effects
such as the gluon azimuthal angular distribution relative to its
parent's momentum direction.

In the respective infrared algebra (QCED) in \eqn{qced}, the average 
Bjorken $x$ values
\begin{eqnarray}
x_{avg}(\rQED)&\cong& \gamma(\rQED)/(1+\gamma(\rQED)),\nonumber\\
x_{avg}(\rQCD)&\cong& \gamma(\rQCD)/(1+\gamma(\rQCD)),
\nonumber
\end{eqnarray}
where
$\gamma(A)=\frac{2\alpha_{A}{\cal C}_A}{\pi}(L_s
-1)$, $A=$ QED, QCD, with ${\cal C}_A=Q_f^2, C_F$, respectively, for 
$A=$ QED, QCD and the big log $L_s$, imply that
QCD dominant corrections happen an
order of magnitude earlier than those for QED. This means 
that the leading $\tilde{\bar\beta}_{0,0}$-level
gives already a good estimate of the size of the interplay between the
higher order QED and QCD effects which we will use to
illustrate \eqn{qced} here.

\section{QED$\otimes$ QCD Threshold Corrections and\\ Shower/ME Matching 
	at the LHC}
The cross section for the processes
$pp\rightarrow V +n(\gamma)+m(g)+X\rightarrow \bar{\ell} \ell'
+n'(\gamma)+m(g)+X$, where $V, \ell, \ell'$ are the vector-boson / lepton
combinations defined in Section 3, 
may be constructed from the parton-level cross section via the usual
formula
(we use the standard notation here~\cite{qced})
\begin{equation}
d\sigma_{\exp} = 
\sum_{i,j}\int dx_idx_j F_i(x_i)F_j(x_j)d\hat\sigma_{\exp}(x_ix_js),
\label{sigtot} 
\end{equation}
In this section, we will use the result in \eqn{qced} here with 
semi-analytical methods and structure functions from Ref.~\cite{mrst1} to
examine the size of QED$\otimes$QCD threshold corrections.
A Monte Carlo realization will appear elsewhere~\cite{elsewh}.

First, we wish to make contact with the
existing literature and standard practice for QCD parton showers
as realized by HERWIG~\cite{herwig} 
and/or PYTHIA~\cite{pythia}. Eventually, we will also make contact with the 
new parton distribution function evolution MC algorithm
in Ref.~\cite{jadskrz}. We intend
to combine our exact YFS-style resummation calculus with HERWIG and/or PYTHIA
by using the latter to generate a parton shower starting from the initial 
$(x_1,x_2)$ point at factorization scale $\mu$, after this point is provided 
by the $\{F_i\}$. This combination of theoretical constructs can be 
systematically improved with
exact fully exclusive results order-by-order in $\alpha_s$, where  
currently the state of the art in such
a calculation is the work in Ref.~\cite{frixw}
which accomplishes the combination of an exact ${\cal O}(\alpha_s)$
correction with HERWIG, where the gluon azimuthal angle 
is averaged in the combination.

The issue of this being an exact rearrangement of the QCD and QED perturbative
series requires some comment. Unlike the threshold resummation techniques
in Refs.~\cite{sterman,cat-tren,berger}, 
we have a resummation which is valid over
the entire phase space. Thus, it is readily applicable to an exact treatment
of the respective phase space in its implementation via MC methods. 

We may illustrate how the combination with PYTHIA/HERWIG may 
proceed as follows. We note that, for example, if we use a quark mass $m_q$ as our collinear limit regulator, DGLAP~\cite{dglap} evolution of the 
structure functions allows us to factorize all the terms that involve powers of the big log $L_c=\ln \mu^2/m_q^2-1$ in such a way that the evolved structure function contains the
effects of summing the leading big logs $L=\ln \mu^2/\mu_0^2$
where the evolution involves initial data at the
scale $\mu_0$. This gives us a result independent of $m_q$ for 
$m_q \downarrow 0$. In the DGLAP theory, the factorization scale
$\mu$ represents the largest $p_{\rm T}$ of the gluon emission included
in the structure function. 

In practice, when we use these structure functions
with an exact result for the residuals in \eqn{qced}, it means that
we must in the residuals omit the contributions from gluon radiation
at scales below $\mu$. This can be shown to amount in most cases to
replacing $L_s=\ln \hat{s}/m_q^2 -1 \rightarrow L_{nls}=\ln\hat{s}/\mu^2$
but in any case it is immediate how to limit the $p_T$ in the 
gluon emission\footnote{ Here, we refer to both on-shell and off-shell
emitted gluons.} so that we do not double count effects.
In other words, we apply the standard QCD factorization of mass singularities
to the cross section in \eqn{qced} in the standard way. We may do it
with either the mass regulator for the collinear singularities or with
dimensional regularization of such singularities. The final result
should be independent of this regulator and this is something that 
we may use as a cross-check on the results. 

This would in practice mean the following: 
We first make an event with the formula
in \eqn{sigtot} which would produce an initial beam state at $(x_1,x_2)$
for the two hard interacting partons at the factorization scale $\mu$
from the structure functions $\{F_j\}$ and a corresponding final state $X$
from the exponentiated cross section in $d\hat\sigma_{\exp}(x_ix_js)$,
where we stress that the latter has had all collinear singularities factorized
so that it is much more convergent then its analog in LEP physics for the
electroweak theory for example.  The standard Les Houches 
procedure~\cite{leshouches} of showering this event $(x_1,x_2,X)$ 
would then be used, employing backward evolution of the
initial partons. If we restrict the $p_T$ as we have
indicated above, there would be no double counting of effects.
Let us call this $p_T$ matching of the shower from the backward
evolution and the matrix elements in the QCED exponentiated cross section.

It is possible, however, to be more accurate
in the use of the exact result in \eqn{qced}. 
Just as the residuals $\tilde{\bar\beta}_{n,m}$$(k_1,\ldots,k_n;$ 
$k'_1,\ldots,k'_m)$ are computed order by order in perturbation 
theory from the corresponding
exact perturbative results by expanding the exponents in \eqn{qced}
and comparing the appropriate corresponding coefficients of the
respective powers of $\alpha^n\alpha_s^m$, so too can the shower
formula which is used to generate the backward evolution be expanded
so that the product of the shower formula's perturbative expansion,
the perturbative expansion of the exponents in \eqn{qced}, and the
perturbative expansions of the residuals can be 
written as an over-all expansion in powers of $\alpha^n\alpha_s^m$
and required to match the respective calculated exact result
for given order. In this way, new shower subtracted residuals, 
$\{\hat{\tilde{\bar\beta}}_{n,m}$$(k_1,\ldots,k_n;$ $k'_1,\ldots,k'_m)\}$, 
are calculated that can be used for the entire gluon $p_{\rm T}$
phase space with an accuracy of the cross section that
should in principle be improved compared with the first procedure for shower
matching presented above. Both approaches are under investigation, where we
note that the shower subtracted residuals have been realized
for the exact ${\cal O}(\alpha)$ luminosity Bhabha process at DAPHNE
energies by the authors in Ref.~\cite{piccinini}.

Returning to the general discussion,
we compute, with and without QED, the ratio
$r_{\exp}=\sigma_{\exp}/\sigma_{\rBorn}$,
where we do not use the narrow resonance approximation, for
we wish to set a paradigm for precision heavy vector boson studies.
The formula which we use for $\sigma_{\rBorn}$ is obtained from that in
\eqn{sigtot} by substituting $d\hat\sigma_{\rBorn}$ for $d\hat\sigma_{\exp}$
therein, where $d\hat\sigma_{\rBorn}$ is the respective parton-level 
Born cross section.  Specifically, we have from \eqn{qcd}
the $\tilde{\bar\beta}_{0,0}$-level result
\begin{equation}
\begin{split}
\hat\sigma_{\exp}(x_1x_2s) = & \int^{v_{\max}}_0 dv\;\gamma_{\rQCED} 
\;v^{\gamma_{\rQCED}-1}F_{\rYFS}(\gamma_{\rQCED})\\
 & \qquad e^{\delta_{\rYFS}}\hat\sigma_{\rBorn}((1-v)x_1x_2s)\\
\end{split}
\end{equation}
where we intend the well-known results for the respective parton-level 
Born cross sections and the value of $v_{\max}$ implied by the experimental cuts
under study. 

What is new here is the value for the QED$\otimes$QCD exponent 
\begin{equation}
\gamma_{\rQCED} = 
	\left(2Q_f^2\frac{\alpha}{\pi}+2C_F\frac{\alpha_s}{\pi}\right)L_{nls}
\label{expnt}
\end{equation}
where $L_{nls}=\ln x_1x_2s/\mu^2$ when $\mu$ is the factorization scale.
The functions $F_{\rYFS}(\gamma_{\rQCED})$ and $\delta_{\rYFS}(\gamma_{\rQCED})$
are well-known~\cite{yfs1} as well:
\begin{equation}
\begin{split}
F_{YFS}(\gamma_{\rQCED})&=\frac{e^{-\gamma_{\rQCED}\gamma_E}}
{\Gamma(1+\gamma_{\rQCED})},\cr
\delta_{YFS}(\gamma_{\rQCED})&=\frac{1}{4}\gamma_{\rQCED} +
\left(Q_f^2\frac{\alpha}{\pi}+C_F\frac{\alpha_s}{\pi}\right)
\left(2\zeta(2)-\frac{1}{2}\right),
\end{split}
\label{yfsfns}
\end{equation}
where $\zeta(2)$ is Riemann's zeta function of argument 2, i.e., $\pi^2/6$,
and $\gamma_E$ is Euler's constant, i.e., 0.5772\ldots .

Using these formulas in \eqn{sigtot} allows us to get the results
\begin{equation}
r_{\exp}=\left\{
\begin{array}{lll}
1.1901&, \qquad \text{QCED}\equiv \text{QCD+QED}, &\text{LHC}\\
1.1872&, \qquad \text{QCD}, &\text{LHC}\\
1.1911&, \qquad \text{QCED}\equiv \text{QCD+QED}, &\text{Tevatron}\\
1.1879&, \qquad \text{QCD}, &\text{Tevatron.}
\end{array}\right.
\label{res1}
\end{equation}
We see that QED is at the level of .3\% at both LHC and FNAL.
This is stable under scale variations~\cite{qced}.
We agree with the results in Refs.~\cite{baurall,ditt,russ,van1,van2}
on both of the respective sizes of the QED and QCD effects.
Furthermore, the QED effect is similar in size to structure function
results found in Refs.~\cite{cern2000,spies,james1,roth,james2}.

\section{Conclusions}
We have shown that YFS theory (EEX and CEEX), when extended to 
non-Abelian gauge theory, allows simultaneous exponentiation of 
QED and QCD, QED$\otimes$QCD exponentiation. For QED$\otimes$QCD we find that
full MC event generator realization is possible in a way that
combines our calculus with HERWIG and PYTHIA in principle.
Semi-analytical results for QED (and QCD) threshold effects agree 
with literature on $Z$ production. As QED is at the .3\% level, 
it is needed for LHC theory predictions at $\lesssim$ 1\%. 
The corresponding analysis of 
the $W$ production is in progress.  We have illustrated
a firm theoretical basis for the realization of the
complete ${\cal O}(\alpha_s^2, \alpha\alpha_s, \alpha^2)$
results needed for the FNAL/LHC/RHIC/ILC physics 
and all of the latter are in progress.

\section*{Acknowledgments}
Work partly supported by US DOE grant DE-FG02-05ER41399 and
by NATO grant PST.CLG.980342.
S.A.Y.\ thanks the organizers of the 2007 Cracow Epiphany Conference
for hospitality.  B.F.L.W.\ thanks Prof.\ W.\ Hollik for the support and kind
hospitality of the MPI, Munich, while a part of this work was
completed. We also thank Prof.\ S.\ Jadach for useful discussions.

\end{document}